\documentclass[%
 reprint,
 superscriptaddress,
 amsmath,amssymb,
 aps,
 prb,
 floatfix,
]{revtex4-2}

\usepackage[none]{hyphenat}
\usepackage{graphicx}
\usepackage{dcolumn}
\usepackage{bm}
\usepackage[utf8]{inputenc}
\usepackage[T1]{fontenc}
\usepackage{mathptmx}
\usepackage{textcomp}
\usepackage{siunitx}
\usepackage{bigints}
\usepackage{upgreek}
\usepackage{hyperref}
\usepackage[switch]{lineno}
\usepackage[table,xcdraw,dvipsnames]{xcolor}
\usepackage{subcaption}
\usepackage{booktabs}
\usepackage{float}
\usepackage{color}
\usepackage{multirow}
\usepackage{placeins}
\usepackage{url}

\newcommand*\diff{\mathop{}\!\mathrm{d}}

\begin{document}

\preprint{APS/123-QED}

\title{Parallel-in-Time Integration of the Landau-Lifshitz-Gilbert Equation with the Parallel Full Approximation Scheme in Space and Time}

\affiliation{Research Platform MMM Mathematics - Magnetism - Materials, University of Vienna, Vienna, Austria}
\affiliation{Physics of Functional Materials, University of Vienna, Vienna, Austria}
\affiliation{Vienna Doctoral School in Physics, University of Vienna, Vienna, Austria}

\author{Robert~Kraft}    
\affiliation{Research Platform MMM Mathematics - Magnetism - Materials, University of Vienna, Vienna, Austria}
\affiliation{Vienna Doctoral School in Physics, University of Vienna, Vienna, Austria}
\author{Sabri~Koraltan}    
\affiliation{Physics of Functional Materials, University of Vienna, Vienna, Austria}
\affiliation{Vienna Doctoral School in Physics, University of Vienna, Vienna, Austria}
\author{Markus~Gattringer}    
\affiliation{Physics of Functional Materials, University of Vienna, Vienna, Austria}
\affiliation{Vienna Doctoral School in Physics, University of Vienna, Vienna, Austria}
\author{Florian~Bruckner}    
\affiliation{Research Platform MMM Mathematics - Magnetism - Materials, University of Vienna, Vienna, Austria}
\affiliation{Physics of Functional Materials, University of Vienna, Vienna, Austria}
\author{Dieter~Suess}    
\affiliation{Research Platform MMM Mathematics - Magnetism - Materials, University of Vienna, Vienna, Austria}
\affiliation{Physics of Functional Materials, University of Vienna, Vienna, Austria}
\author{Claas~Abert}    
\affiliation{Research Platform MMM Mathematics - Magnetism - Materials, University of Vienna, Vienna, Austria}
\affiliation{Physics of Functional Materials, University of Vienna, Vienna, Austria}
\date{\today}

\begin{abstract}
Speeding up computationally expensive problems, such as numerical simulations of large micromagnetic systems, requires efficient use of parallel computing infrastructures. While parallelism across space is commonly exploited in micromagnetics, this strategy performs poorly once a minimum number of degrees of freedom per core is reached. We use magnum.pi, a finite-element micromagnetic simulation software, to investigate the Parallel Full Approximation Scheme in Space and Time (PFASST) as a space- and time-parallel solver for the Landau-Lifshitz-Gilbert equation (LLG). Numerical experiments show that PFASST enables efficient parallel-in-time integration of the LLG, significantly improving the speedup gained from using a given number of cores as well as allowing the code to scale beyond spatial limits.
\end{abstract}

\maketitle


\section{Introduction}
\label{sec:introduction}

Computationally expensive simulations of large micromagnetic systems can be significantly accelerated by carrying out computations in parallel. Parallelism in space via distributed meshes or grids is commonly exploited in micromagnetic simulation codes, for instance in CPU software such as \texttt{OOMMF} \cite{Donahue1999}, \texttt{mumax} \cite{Vansteenkiste2014} or GPU codes like \texttt{magnum.np} \cite{Bruckner2023} or \texttt{BORIS} \cite{Lepadatu2023}. However, the efficiency of this strategy drops sharply when the number of degrees of freedom per core gets too small, as the communication overhead begins to dominate the computational cost. Therefore, alternative avenues for parallelism such as computing multiple timesteps in parallel must be considered. This has already been attempted almost 60 years ago \cite{Nievergelt1964} but only recently gained greatly increased interest. As a result, a large variety of so-called parallel-in-time integration algorithms mostly based on Parareal \cite{Lions2001} have been developed in the last decade \cite{Gander2015}, such as Multigrid Reduction In Time (MGRIT) \cite{Friedhoff2012} or the Parallel Full Approximation Scheme in Space and Time (PFASST) \cite{Emmett2012}, to name a few. The aforementioned algorithms have been applied successfully to a variety of problems \cite{Fischer2005,Legoll2022,Schoeps2018,Speck2014,Goetschel2018}, enhancing parallel efficiency and allowing codes to scale much further than with spatial parallelism alone. Nonetheless, parallel-in-time integration has not yet been adapted to micromagnetics. In this work, we therefore investigate parallel-in-time integration of the Landau-Lifshitz-Gilbert equation (LLG) with the PFASST algorithm to gain an understanding of the potential benefits and difficulties. After explaining the numerical prerequisites and implementation, numerical experiments giving a first impression on the performance of the scheme are presented and discussed.

\section{Methods}
\label{sec:methods}
In the following, basic concepts of micromagnetism and the prerequisites for the implemented PFASST scheme will be briefly explained, followed by a description of the algorithm itself. For a more comprehensive overview of micromagnetics refer to \cite{Abert2019} and for more information regarding parallel-in-time integration, consider visiting the \href{http://parallel-in-time.org/}{parallel-in-time.org} website.

\subsection{Micromagnetics and the LLG}
\label{subsec:basics}
The micromagnetic model strikes a balance between accuracy and efficiency by assuming that the quantum mechanical exchange interaction between neighboring spins is strong enough to align magnetic moments in parallel at distances much greater than the underlying lattice constant, a characteristic property of ferromagnets. As a result, the magnetization of a material can be written as a continuous, vector-valued function $\boldsymbol{M}(\boldsymbol{x})=M_\text{s}\, \boldsymbol{m}(\boldsymbol{x})$ where $\left\lVert\boldsymbol{m}(\boldsymbol{x})\right\rVert = 1$ and $M_\text{s}$ refers to the constant saturation magnetization. Energy contributions influencing the magnetization are then expressed as integrals involving the magnetization function. Notable contributions to the total energy $E$ are for example the Zeeman energy originating from an external field, the demagnetizing energy resulting from the field caused by the material itself, the aforementioned exchange energy, or effects such as anisotropy. Furthermore, $E$ is used to define the so-called effective field $\boldsymbol{H}^{\text{eff}}$,
\begin{equation}
    \boldsymbol{H}^{\text{eff}}=-\frac{1}{\mu_0\,M_\text{s}}\frac{\updelta E}{\updelta m}\,,
\end{equation}
where $\mu_0\approx\SI{1.2566e-6}{\newton\per\ampere^2}$ refers to the vacuum permeability and $\updelta$ denotes the functional derivative. While the total energy can be minimized directly to find equilibrium configurations, the importance of the effective field lies in modeling magnetization dynamics when inserted into the LLG,
\begin{equation}
    \frac{\partial \boldsymbol{m}}{\partial t} = -\frac{\gamma}{1 + \alpha^2}\,\boldsymbol{m}\times\boldsymbol{H}^{\text{eff}}-\frac{\alpha\,\gamma}{1 + \alpha^2}\,\boldsymbol{m}\times(\boldsymbol{m}\times\boldsymbol{H}^{\text{eff}})\,,
    \label{eq:LLG_expl}
\end{equation}
where $\alpha$ refers to a dimensionless damping parameter modeling the relaxation of the magnetization toward the field lines, whereas the reduced gyromagnetic ratio $\gamma = \gamma_e/\mu_0 \approx \SI{2.2128e5}{\metre\per{\ampere\second}}$ determines the frequency of the precession around the field lines. Spatial discretization of the LLG is commonly done with either finite-difference (FDM) \cite{Donahue1999,Vansteenkiste2014,Bruckner2023,Lepadatu2023} or finite-element (FEM) \cite{Abert2013, Chang2011, TetraX} methods. In this work, the finite-element micromagnetic solver \texttt{magnum.pi} \cite{Abert2013} is used. It is based on \texttt{firedrake} \cite{Rathgeber2016}, which enables distributing meshes and therefore spatial parallelism for finite-element discretizations of partial differential equations. By default, time integration in \texttt{magnum.pi} is done with the \texttt{SUNDIALS} \texttt{CVODE} \cite{Hindmarsh2005} module. More specifically, a preconditioned, adaptive second-order Backward Differentiation Formula (BDF) timestepper described in \cite{Suess2002} is employed. This fully implicit time integration method is suited well for dealing with the high stiffness introduced by the exchange interaction.

\subsection{Spectral Deferred Corrections}
\label{subsec:sdc}
The fundamental ingredient of PFASST is a class of timesteppers referred to as Spectral Deferred Correction (SDC) methods \cite{Dutt2000}. Such timesteppers solve ordinary differential equations by iteratively correcting an initial approximation using only low order methods. For the sake of brevity, the method is commonly derived using a preconditioned Richardson iteration viewpoint \cite{Ruprecht2016,Huang2006} as follows:
Consider an initial value problem in integral form,
\begin{equation}
    u(t) = u_\text{a} + \int_{t_\text{a}}^{t} f\big(\tau, u(\tau)\big)\,\diff\tau\,,~ u(t_\text{a}) = u_\text{a}\,,
    \label{eq:integralform}
\end{equation}
to be integrated until time $t_\text{b}$. Constructing an SDC scheme carrying out this integration begins with the choice of outer quadrature nodes ($t_1,\dots,t_m$) on the interval $[t_\text{a},t_\text{b}]$ according to some quadrature rule (e.g. Gauss-Legendre). Then, integrals of the right-hand side $f$ to these outer quadrature nodes are approximated with the integration matrix $\boldsymbol{Q} \in \mathbb{R}^{m\times m}$:
\begin{equation}
    \int_{t_\text{a}}^{t_i} f\big(\tau,u(\tau)\big)\diff\tau \approx \sum_{j=1}^m Q_{ij}\, f\big(t_j,u(t_j)\big)\,, \quad i = 1,\dots m
    \label{eq:approxint}
\end{equation}
In other words, row $i$ of matrix $\boldsymbol{Q}$ contains quadrature weights that integrate from $t_\text{a}$ to $t_i$. Obtaining said weights usually involves an interpolation step \cite{Bolten2017}. Inserting the approximations \eqref{eq:approxint} into \eqref{eq:integralform} gives rise to the so-called collocation problem
\begin{equation}
    (\boldsymbol{I}_m - \boldsymbol{Q}\boldsymbol{F})[\boldsymbol{u}] = \boldsymbol{u}_\text{a}\,,
    \label{eq:collocation_problem}
\end{equation}
which, when solved, gives an approximation to the solution at times $t_i$, $\boldsymbol{u} = (u_1,\dots,u_m)$ where $u_i=u(t_i)$. Here, $\boldsymbol{u}_\text{a} = (u_\text{a},\dots,u_\text{a}) \in \mathbb{R}^m$ refers to a vector filled with the initial value, $\boldsymbol{I}_m$ to the $m$-dimensional identity operator and $\boldsymbol{F}[\boldsymbol{u}]=(f(t_1,u_1),\dots,f(t_m,u_m))\in \mathbb{R}^m$ evaluates the right-hand side at the outer quadrature nodes. SDC methods now solve the nonlinear equation \eqref{eq:collocation_problem} by applying Richardson iteration with a preconditioner $\boldsymbol{Q}^\text{P}$:
\begin{equation}
    (\boldsymbol{I}_m - \boldsymbol{Q}^{\text{P}}\boldsymbol{F})[\boldsymbol{u}^{k+1}] = (\boldsymbol{Q} - \boldsymbol{Q}^{\text{P}})\boldsymbol{F}[\boldsymbol{u}^k] + \boldsymbol{u}_\text{a}
    \label{eq:sdc_sweep}
\end{equation}
The preconditioner $\boldsymbol{Q}^\text{P}$ usually is a simpler, lower triangular integration rule for the subintervals, e.g. backward integration. If the preconditioner is strictly lower triangular (e.g. forward integration), solving a (non)linear system for the iteration is not required and the scheme is explicit. Computing one iteration of \eqref{eq:sdc_sweep} is often called an SDC sweep. Carrying out multiple sweeps sequentially improves the accuracy of the approximation. In fact, one SDC sweep raises the order of accuracy in the timestep by one or more (depending on the accuracy of the preconditioner), up until the order of the outer quadrature \cite{Causley2019}. Furthermore, using two separate preconditioners, one of which lower triangular ($\boldsymbol{Q}^{\text{P}}_{\text{im}}$) and one strictly lower triangular ($\boldsymbol{Q}^{\text{P}}_{\text{ex}}$) allows for constructing an implicit-explicit splitting (IMEX) version of SDC. With a correspondingly split right-hand side $\boldsymbol{F} = \boldsymbol{F}_\text{ex} + \boldsymbol{F}_\text{im}$, the sweeps then look as follows:
\begin{multline}
    (\boldsymbol{I}_m - \boldsymbol{Q}^{\text{P}}_{\text{im}}\,\boldsymbol{F}_{\text{im}} - \boldsymbol{Q}^{\text{P}}_{\text{ex}}\,\boldsymbol{F}_{\text{ex}})[\boldsymbol{u}^{k+1}] = \\
    (\boldsymbol{Q}\boldsymbol{F} - \boldsymbol{Q}^{\text{P}}_{\text{im}}\,\boldsymbol{F}_{\text{im}} - \boldsymbol{Q}^{\text{P}}_{\text{ex}}\,\boldsymbol{F}_{\text{ex}})[\boldsymbol{u}^k] + \boldsymbol{u}_\text{a}
    \label{eq:sdc_imex} 
\end{multline}

\subsection{Multilevel SDC}
\label{sec:mlsdc}
Multigrid methods outsource computationally expensive tasks to a coarser discretization where they can be carried out in a cheap manner. This can be used to accelerate convergence for SDC timesteppers and is used in PFASST to facilitate a cheap estimation of the solution at future timesteps. In so-called Multilevel SDC (MLSDC) methods, the Full Approximation Scheme (FAS) multigrid correction technique \cite{Henson2002} is used to obtain a correction for the right-hand side integrals at the coarse level,
\begin{equation}
    \boldsymbol{\tau} = \boldsymbol{T}_{\text{f}}^{\text{c}}\,\boldsymbol{Q}\,\boldsymbol{F}[\boldsymbol{u}^k] - \tilde{\boldsymbol{Q}}\,\boldsymbol{T}_{\text{f}}^{\text{c}}\,\boldsymbol{F}[\boldsymbol{u}^k]\,,
    \label{eq:taucorr}
\end{equation}
where $\boldsymbol{T}_{\text{f}}^{\text{c}}$ refers to a restriction operator which can transform onto both a coarser mesh and time discretization (i.e. fewer nodes in the outer quadrature). Furthermore, $\tilde{\boldsymbol{Q}}$ represents the integration matrix on the coarse level. Adding $\boldsymbol{\tau}$ to the right-hand side of \eqref{eq:sdc_sweep} for the coarse level SDC sweeps enhances accuracy, often described as computing the solution at the resolution of the coarse grid but with the accuracy of the fine grid. Overall, a complete MLSDC iteration involves the following:
\begin{enumerate}
    \item Restrict the fine level approximation: $\tilde{\boldsymbol{u}}^k = \boldsymbol{T}_{\text{f}}^{\text{c}}[\boldsymbol{u}^k]$.
    \item Compute $\boldsymbol{\tau}$ according to \eqref{eq:taucorr} and carry out corrected SDC sweeps on the coarse level to obtain $\tilde{\boldsymbol{u}}^{k+1/2}$.
    \item Get a coarse level correction for the solution: $\tilde{\boldsymbol{\delta}}^k=\tilde{\boldsymbol{u}}^{k+1/2}-\tilde{\boldsymbol{u}}^{k}$.
    \item Prolong the correction to the fine level with the prolongation operator $\boldsymbol{T}_{\text{c}}^{\text{f}}$, $\boldsymbol{\delta}^k=\boldsymbol{T}_{\text{c}}^{\text{f}}[\tilde{\boldsymbol{\delta}}^k] $, and apply it to the fine level approximation: $\boldsymbol{u}^{k+1/2} = \boldsymbol{u}^k + \boldsymbol{\delta}^k$.
    \item Carry out fine level SDC sweeps to obtain $\boldsymbol{u}^{k+1}$.
\end{enumerate}
Aside from the two-level example above, this can also be carried out on three or more levels \cite{Emmett2012, Bolten2017}.

\subsection{Parallel Full Approximation Scheme in Space and Time}
\label{sec:pfasst}
\begin{figure}[]
    \centering
    \includegraphics[width=\columnwidth]{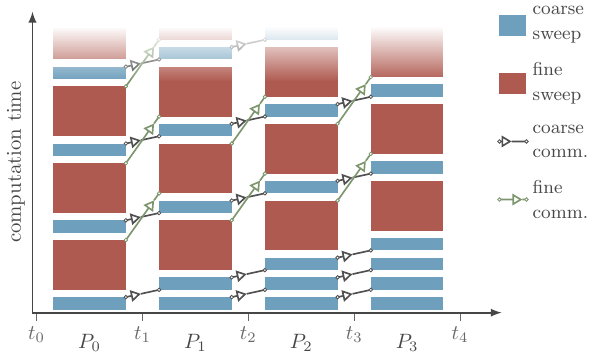}
    \caption{Two-level PFASST iterations on four processes visualized with \texttt{pfasst-tikz} \protect{\cite{Koehler2015}}. End results of coarse (blue) and fine (red) sweeps are sent forward to the next process (arrows), where they are used as initial values.}
    \label{fig:pfasst_comm}
\end{figure}

The structure of the MLSDC algorithm shown above is suited well for introducing parallelism in time. All that remains to do is strategically passing intermediate results forward, to be used as initial values for processes that calculate future timestep values. Consider Fig.\ \ref{fig:pfasst_comm}, which depicts two-level MLSDC iterations that take place on multiple processes ($P_0, \dots,  P_3$), computing multiple timesteps ($[t_0,t_1], [t_1,t_2], \dots $) concurrently. Aside from an initialization procedure at the beginning, the only additions to the MLSDC algorithm are passing the coarse level result forward to the next process upon completing coarse sweeps and passing fine level results forward to the next iteration on the next process. Regarding the initialization procedure, the repeated coarse sweeps shown in the Fig.\ are a common choice.

There are many variants of PFASST with different communication patterns or other modifications. The \texttt{pySDC} library \cite{Speck2019} for example, which we used to prototype the PFASST and SDC solvers implemented in this work, adopts a multigrid viewpoint. While this enables a more general implementation of the algorithm, it also decreases efficiency. Here however we use the classical two-level variant depicted in Fig.\ \ref{fig:pfasst_comm}. When compared to serial SDC, the parallel efficiency of this variant is at most $K_\text{s}/K_\text{p}$ \cite{Minion2011}, where $K_\text{s}$ refers to the number of iterations required to solve the problem with SDC, whereas $K_\text{p}$ refers to the required PFASST iterations to solve the problem with the same accuracy. This is an improvement over the upper efficiency bound of $1/K$ for Parareal \cite{Lions2001}, where $K$ is the number of iterations required for convergence.

\subsection{Implementation}
\label{sec:implementation}
The first of many choices in implementing a PFASST scheme is choosing a numerical quadrature for the SDC sweeps. Usually, quadrature types with the highest possible order given a certain number of nodes are chosen. Gauss-Legendre quadrature for instance integrates polynomials of up to degree $2\,n-1$ exactly with $n$ nodes and is optimal in this sense. On the other hand, Gauss-Legendre nodes on a given interval do not include the endpoint of said interval, where the desired solution might lie. Hence, an additional routine for calculating the endpoint is needed to integrate ODEs with such a quadrature type. This issue can be avoided entirely with quadratures that include the endpoint, such as Gauss-Radau quadrature which integrates polynomials of up to degree $2\,n-2$ exactly with $n$ nodes. In this work, both Gauss-Legendre and Gauss-Radau nodes are used, where the solution at the endpoint is computed by integrating with the outer quadrature weights if necessary.

Next, the preconditioners must be assembled. Common choices here are forward integration as an explicit preconditioner and the so-called \textit{LU-trick} \cite{Weiser2015},
\begin{equation}
    \boldsymbol{Q}^{\text{P}}_{\text{LU}} = \boldsymbol{U}^\text{T}\,, \text{ where } \boldsymbol{Q}^\text{T} = \boldsymbol{L}\,\boldsymbol{U}\,,
    \label{eq:LU_trick} 
\end{equation}
as an implicit preconditioner which performs favorably for stiff problems. Both preconditioning methods are used in the code, because the LLG is not only stiff, but also a good candidate for implicit-explicit time integration: On the one hand, the stiffness is mainly introduced by the short-ranged exchange interaction, which is relatively cheap to compute and can thus be integrated implicitly without issue. On the other hand, the LLG permits a very straightforward splitting of the right-hand side into the contributions of individual field terms, as it is linear in the effective field. In the implemented code, users can choose which field terms are integrated explicitly or implicitly. One major benefit of using such an IMEX time integrator for the LLG is that repeated evaluations of the expensive demagnetizing field can be avoided. Therefore, we always use the IMEX variant of SDC in the numerical experiments of section \ref{sec:results}, where only the exchange field is evaluated implicitly.

The nonlinear system that arises in the SDC sweeps is solved with line search Newton iteration, more specifically the \texttt{PETSc} \cite{balay2019petsc} implementation with the Flexible Generalized Minimum Residual Method (FGMRES) as the linear solver. The necessary right-hand side and Jacobian evaluation routines are already part of \texttt{magnum.pi}. Together with \texttt{firedrake}, this enables spatial parallelism.

Finally, transfer operators between the coarse and fine space/time discretizations must be implemented. In this work, the \texttt{firedrake.multigrid} module is used for spatial transfers, as it allows for restriction and prolongation between distributed meshes. For the temporal domain, restriction and prolongation are realized with interpolation between the coarse and fine level quadrature nodes. 

Regarding error control, serial SDC timesteppers have access to a very simple error estimate: Since the order of accuracy of the solution increases with every sweep, subtracting the solution values of the previous and current iteration gives an estimate for the local truncation error of the previous iterate. The implemented stepsize adaptation is then straightforward: If the error tolerance is not met, the stepsize is decreased by $50\%$, whereas the stepsize is increased by $25\%$ after 10 accepted steps. Additionally, as long as the error exceeds $80\%$ of the tolerance, the stepsize is not increased further. As an alternative to the error estimate, the maximum residual of the collocation problem \eqref{eq:collocation_problem} can be considered for error control. This is particularly useful for comparing the accuracy of SDC and PFASST results.

Lastly, it is worth mentioning that the implemented PFASST controller is very similar in structure to the classical two-level variant in Fig.\ \ref{fig:pfasst_comm}, with some modifications for increased efficiency. For instance, reevaluating function values when returning from the coarse to the fine grid can be avoided by also correcting the function values in the same way as the approximate solution. Also, CPU idle times are avoided as much as possible by immediately starting computation on a new timestep once a timestep has converged, instead of waiting for a complete set of timesteps to converge and restarting the PFASST algorithm to compute the next set.

\section{Numerical Experiments}
\label{sec:results}
In the following, we demonstrate the efficiency of the implemented timestepping methods by simulating various micromagnetic processes. Exploring not only novel parallel timestepping methods but also new kinds of High Performance Computing (HPC) infrastructure, simulations are carried out on the Google Cloud. Here, the \texttt{HPC toolkit} \cite{Platform2023} is used to first build a \texttt{magnum.pi} Virtual Machine (VM) image and then create single VM instances or even entire \texttt{SLURM} \cite{Yoo2003} clusters on the cloud. In this work, the \texttt{c2d} machine type is used for all experiments. More specifically, unless stated otherwise simulations are carried out on \texttt{c2d-highcpu-32} nodes with 16 physical processor cores each, where the compact placement rule for better inter-node communication has been enabled. For another example of micromagnetics code running on cloud services, consider \cite{Jermain2016}.

\subsection{Standard Problem 4}
\label{subsec:sp4_sdc}
PFASST relies entirely on SDC methods for timestepping. As such, investigating the performance of serial SDC timestepping methods gives valuable insight about the potential performance of a PFASST scheme. This preliminary experiment examines SDC as a serial timestepping method for the LLG by comparing it to the BDF method normally employed in \texttt{magnum.pi}. Additionally, a first small-scale time-parallel run is attempted. As parallelism in space is not of interest here, we simulate the well-known $\mu$MAG Standard Problem \#4  (SP4) \cite{StandardProblem4}.

\begin{figure}[]
    \centering
    \includegraphics[width=\columnwidth]{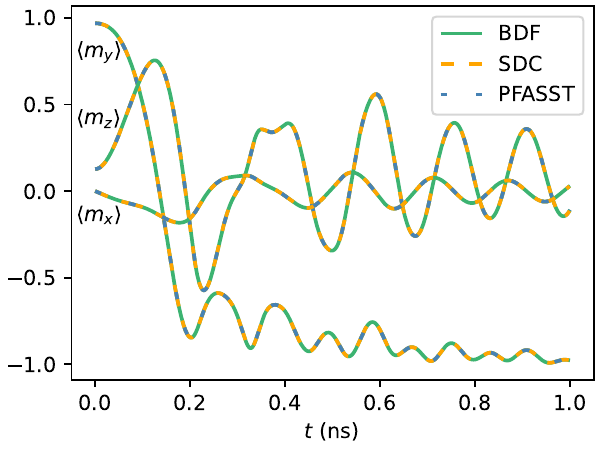}
    \caption{Average magnetization in the coordinate directions for SP4. The BDF run with $tol=\num{e-10}$, SDC run with $tol=\num{e-5}$ and PFASST run with $n_\text{t}=3$ are shown.}
    \label{fig:sp4_results}
\end{figure}

First, we simulate SP4 with \texttt{magnum.pi}'s built-in adaptive BDF solver and the adaptive SDC timestepper at varying tolerances. These runs are compared to a reference solution obtained with BDF at very strict ($tol$ = \num{e-10}) error tolerances (Tab.\ \ref{tab:sdctimings}, Fig.\ \ref{fig:sp4_results}).

\begin{table}[]
    \normalsize\centering
    \caption{Tolerances $tol$, runtimes $t$, average relative $L^2$-norm errors $\overline{e}_\text{rel}$ and number of external $n_\text{ext}$, demagnetizing $n_\text{dem}$ as well as exchange $n_\text{exc}$ field evaluations for SP4.}
    \begin{tabular}{ccccccc}
    \toprule
                  & $tol$ & $t$ (s) & $\overline{e}_\text{rel}$ & $n_\text{ext}$ & $n_\text{dem}$ & $n_\text{exc}$ \\ \midrule
    SDC    & \num{1.0e-2} & \num{47}&  \num{1.54e-3}   &  \multicolumn{2}{c}{2324}       &   6426     \\
    BDF    & \num{1.0e-5} & \num{66}&  \num{3.42e-3}   &  \multicolumn{3}{c}{5124}      \\ 
    SDC    & \num{1.0e-5}  & \num{140}&  \num{9.66e-6}   &   \multicolumn{2}{c}{6111}      &   27798     \\
    BDF    &  \num{2.5e-9} &\num{708} & \num{1.20e-5} &  \multicolumn{3}{c}{67587}     \\ \bottomrule
    \end{tabular}
    \label{tab:sdctimings}
\end{table}

On average, the SDC runs are slightly more accurate than comparable BDF runs but also significantly faster, especially at high accuracy. The speedup is explained as follows: First, a large fraction of the computationally costly stray field evaluations is avoided by employing the IMEX variant of SDC instead of the fully implicit BDF method. Additionally, two Gauss-Legendre nodes were used for the high-accuracy SDC run. This results in a third order timestepping method as opposed to the second order BDF method.

\begin{figure}[]
    \centering
    \includegraphics[width=\columnwidth]{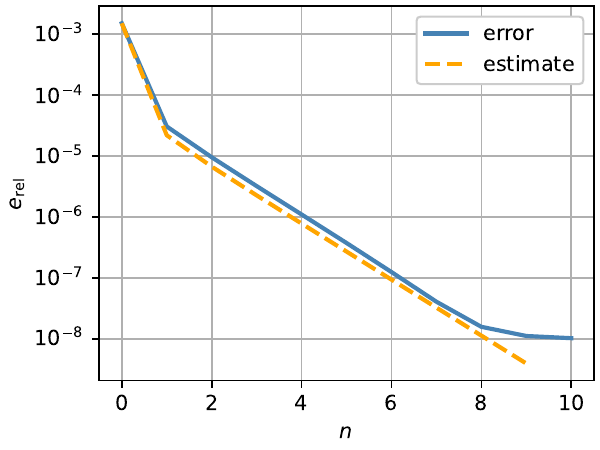}
    \caption{Relative $L^2$-norm error $e_\text{rel}$ and error estimate at SDC iteration $n$ on the first timestep of SP4.}
    \label{fig:sp4_timestep_err}
\end{figure}

Next, we determine the error of individual SDC iterations on the first timestep of SP4. To this end, we use five Gauss-Legendre nodes for SDC and a BDF reference solution with even stricter tolerances ($tol = \num{e-14}$). Plotting the errors obtained this way against the iteration number confirms the theoretically predicted exponential convergence (Fig.\ \ref{fig:sp4_timestep_err}). Also, one can clearly see how convergence stops after the accuracy of the quadrature is exhausted and how the error estimate gives a good approximation of the error until then.

Finally, since the PFASST controller does not yet support adaptive stepsizes, we use a fixed stepsize of $dt = \SI{e-12}{\second}$ and a relative residual tolerance of $restol = \num{e-6}$ to compute SP4 parallel-in-time on up to three cores (Tab.\ \ref{tab:sp4_pfasst_timings}). Here, the fine level mesh is the same as in the previous experiments with 15000 elements in total, whereas the coarse level mesh consists of only 600 elements. The residual tolerance is chosen such that the results roughly match the accuracy of the SDC/BDF runs in Tab.\ \ref{tab:sdctimings}.
\begin{table}[]
    \centering\normalsize
    \caption{Number of cores $n_\text{t}$, runtimes $t$, speedups $S$ and average relative $L^2$-norm errors $\overline{e}_\text{rel}$ when computing SP4 with PFASST.}
    \begin{tabular}{ccccc}
    \toprule
    & $n_\text{t} $& $t$ (s) & $S$ & $\overline{e}_\text{rel}$ \\ \midrule
    \multirow{3}{*}{PFASST}& 1&\num{246} &  1.00   &  \num{3.87e-6}       \\ 
    & 2&\num{148} & 1.66    &  \num{5.29e-6}     \\ 
    & 3&\num{114} &  2.16   &  \num{1.27e-5}     \\ \bottomrule
    \end{tabular}
    \label{tab:sp4_pfasst_timings}
\end{table}
As can be seen in Tab.\ \ref{tab:sp4_pfasst_timings}, using PFASST on a single core (i.e. MLSDC) is initially slower than using adaptive SDC. This relatively large slowdown is primarily caused by the already very small fine mesh size, which makes the additional overhead introduced in PFASST much more noticeable and also diminishes the speedup gained from transferring to the coarse level. Despite this, a parallel speedup of up to 2.16 is achieved with three cores in time. Note that the average relative error increases with the number of cores in time. However, this is expected and originates from the fact that one single PFASST iteration can significantly reduce the relative residual. Therefore, while all results meet the same residual tolerance, they vary in accuracy depending on how much the relative residual decreases below the tolerance on the final iteration of every timestep. It is also worth mentioning that time parallelism on more than a few cores is likely not very efficient for problems like SP4, unless very high accuracies are desired. This is due to the fact that the magnetization changes quickly in SP4, making future timesteps hard to predict. Consequently, parallel efficiency decreases disproportionately as more and more cores are used and the time interval that is computed in parallel grows. Nonetheless, for the specified residual tolerance the time-parallel PFASST runs already offer a small speedup over serial runs.

\subsection{Airbox Hysteresis}
\label{sec:airbox_hyst}

\begin{figure}[]
    \centering
    \includegraphics[width=\columnwidth]{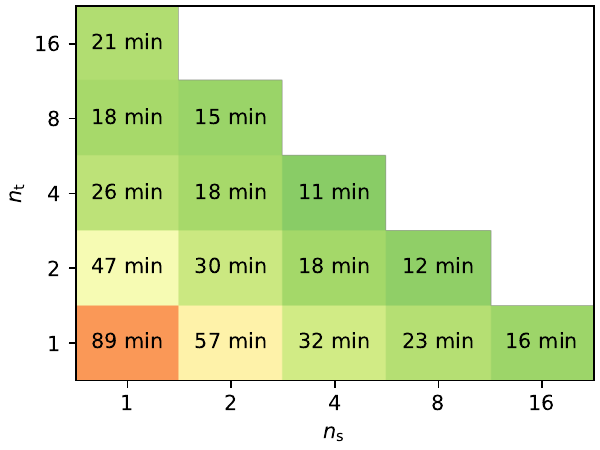}
    \caption{PFASST timings for the hysteresis problem, solved with $n_\text{s}$ cores in space and $n_\text{t}$ parallel timesteps on a single node.}
    \label{fig:airbox_rel_timings}
\end{figure}

\begin{figure}[]
    \centering
    \includegraphics[width=\columnwidth]{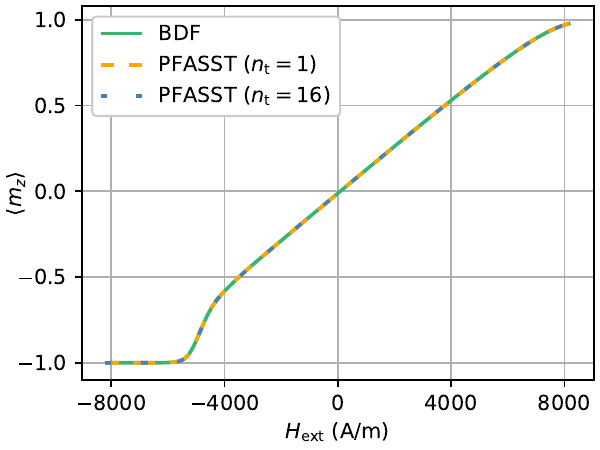}
    \caption{Hysteresis curves computed with BDF and PFASST, where $n_\text{t}$ time intervals are computed in parallel and $H_\text{ext}$ refers to the $z$-component of the external field.}
    \label{fig:airbox_results}
\end{figure}

Unlike the switching process in the previous subsection, magnetization generally changes slowly when computing hysteresis loops. This should improve parallel efficiency and allow using more cores for time parallelism. Therefore, we investigate this problem type next: We set a $\SI{1000}{\nano\metre} \times \SI{1000}{\nano\metre}\times \SI{20}{\nano\metre}$ rectangular domain to a constant initial magnetization of $\boldsymbol{m} = (0.01,0,-1)^\text{T}$ before normalization and then reverse the magnetization by simulating the LLG for $t_{\text{end}} = \SI{2}{\nano\second}$ with an external field of the form
\begin{equation*}
    \boldsymbol{H}_{\text{ext}}(t) = \left(\frac{2t}{t_{\text{end}}}-1\right)\, \frac{B_{\text{max}}}{\mu_0}\,\boldsymbol{e}_{z}\,,~~B_{\text{max}} = \SI{1}{\tesla}\,,
\end{equation*}
where $\boldsymbol{e}_{\text{z}}$ refers to the unit vector in $z$-direction. The material parameters are the same as in SP4, except for the damping where we set $\alpha=1$. In other words, we compute one branch of a hysteresis loop in the direction of the hard axis. Because the hybrid finite-element-boundary-element (FEM-BEM) demagnetizing field solver in \texttt{magnum.pi} does not yet support spatial parallelism, this is done with the airbox solver module and a surrounding airbox of the dimensions $\SI{4000}{\nano\metre} \times \SI{4000}{\nano\metre}\times \SI{400}{\nano\metre}$. We use a mesh with 33033 elements on the coarse PFASST level and, after refinement by bisection, 264264 elements on the fine level. With seven Gauss-Radau nodes on the fine and coarse level, a fixed stepsize of $dt=\SI{2e-11}{\second}$ and a relative residual tolerance of $restol=\num{e-8}$, we obtain the runtimes in Fig.\ \ref{fig:airbox_rel_timings} on a single node. To ensure correctness, we also carry out a BDF run with $tol=\num{e-7}$ and plot the hysteresis curve against selected PFASST results (Fig.\ \ref{fig:airbox_results}).

Notably, up until eight cores, pure time parallelism is more efficient than pure spatial parallelism. This is most likely a consequence of the still relatively small mesh used in the simulations. On the other hand, the efficiency of the time-parallel approach drops quite quickly, such that using 16 cores in time actually slows the computation down compared to using only eight cores.

\begin{figure}[]
    \centering
    \includegraphics[width=\columnwidth]{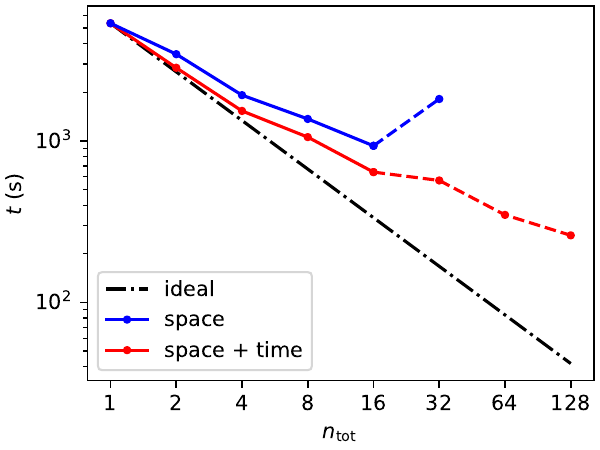}
    \caption{Strong scaling runtimes $t$ for the hysteresis problem, utilizing $n_\text{tot}$ cores on individual nodes (solid lines) and across several nodes (dashed lines). The ideal scaling (dashed-dotted line) is inversely proportional to $n_\text{tot}$.}
    \label{fig:airbox_strsc}
\end{figure}

Finally, Fig.\ \ref{fig:airbox_strsc} compares the scaling of space-parallel only runs with space- and time-parallel runs on one (solid lines) or several (dashed lines) nodes.
Overall, by using 128 cores in total the runtime could be reduced by a factor of 20.56, where the speedup of 5.75 with spatial parallelism alone is enhanced by an additional factor of 3.58 through time parallelism. One should however note that the residual tolerance used in this numerical experiment is relatively strict for simulating such a simple problem, which increases the parallel efficiency of PFASST. On top of that, the coarse level resolution is already quite good, which similarly benefits PFASST. In any case, the results show that, unlike spatial parallelism, parallel-in-time integration can be used efficiently across multiple nodes. This observation is especially interesting in the context of micromagnetic GPU codes, where time parallelism could offer a straightforward way to implement multi-GPU extensions.

\subsection{Magnonic damping}

\begin{figure}[]
    \centering
    \includegraphics[width=\columnwidth]{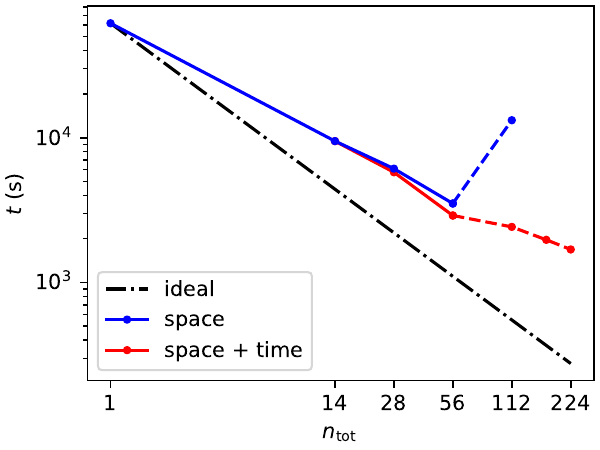}
    \caption{Strong scaling runtimes $t$ for the magnon damping problem, utilizing $n_\text{tot}$ cores on individual nodes (solid lines) and across several nodes (dashed lines). The ideal scaling (dashed-dotted line) is inversely proportional to $n_\text{tot}$.}
    \label{fig:magnon_strsc}
\end{figure}

Finally, we consider a problem of oscillatory nature, where high accuracies are often desired. In \cite{Gattringer2022}, a spin pumping model for \texttt{magnum.pi} based on spin-diffusion is implemented and used to simulate the propagation of a magnon damped by spin torque. More specifically, the magnon is driven by an oscillating magnetic field on one side of a $\SI{1251}{\nano\metre}\times\SI{10}{\nano\metre}\times\SI{10}{\nano\metre}$ waveguide with uniaxial anisotropy along the $y$-axis and damped by a spin sink placed on the waveguide starting from $y=\SI{251}{\nano\metre}$. Furthermore, the simulation is carried out with zero damping ($\alpha=0$) and with the demagnetizing field disabled. We simulate this magnonic damping process for \SI{3}{\nano\second} with the same waveguide structure and material parameters. As for the solver parameters, the PFASST runs utilize a coarse mesh with 84436 elements and a fine mesh resulting from bisection with 675488 elements alongside three Gauss-Radau nodes, a fixed stepsize of $dt = \SI{5e-12}{s}$ as well as a relative residual tolerance of $rtol = \num{e-5}$. Because the spin torque calculations are relatively expensive on a mesh of this size, we now use the \texttt{c2d-highcpu-112} machine type on the Google Cloud, with 56 physical cores per node. Once again, we first simulate on a single node and then extend the parallelism by computing timesteps in parallel on up to four nodes, for a total maximum of 224 cores. This results in the strong scaling depicted in Fig.\ \ref{fig:magnon_strsc}.

\begin{figure*}[]
    \centering
    \includegraphics[width=\textwidth]{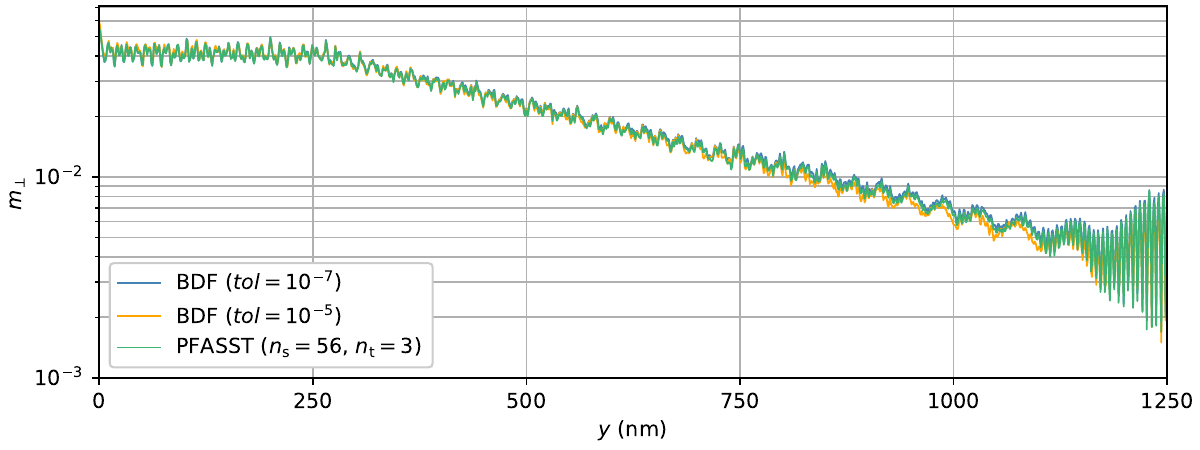}
    \caption{Transversal magnetization component magnitude along the $y$ axis $m_\perp$  at $t = \SI{2.4}{\nano\second}$ in the magnon damping simulations for selected time integration methods and tolerances.}
    \label{fig:magnon_results}
\end{figure*}

At 56 cores, a speedup of 17.56 is achieved with pure spatial parallelism and an additional speedup of 2.08 when using four such nodes parallel-in-time for a total speedup of 36.51. As comparing the performance of PFASST to BDF is also interesting here, we compute a BDF solution with similar accuracy. For this experiment, the quantity of interest is the magnitude of the transversal magnetization component along the $y$-axis, $m_\perp$, at $t = \SI{2.4}{\nano\second}$ (Fig.\ \ref{fig:magnon_results}). Therefore, we use this quantity to estimate the accuracies of our solutions by comparing to a reference BDF run with $tol = \num{e-7}$ (Tab.\ \ref{tab:magnon_timings}).

\begin{table}[]
    \centering\normalsize
    \caption{Number of cores used in space $n_\text{s}$, parallel timesteps $n_\text{t}$, runtimes $t$ and average relative errors $\overline{e}_\text{rel}$ of $m_\perp$ for the magnon damping problem.}
    \begin{tabular}{lcccc}
    \toprule
                                        & $n_\text{s}$ & $n_\text{t}$ & $t$ (h) & $\overline{e}_\text{rel}$ \\ \midrule
    BDF ($tol=\num{e-5}$)                    & 1    & 1    & 36.11           & \num{6.12e-04}   \\ \midrule
    \multirow{10}{*}{PFASST} & 1    & 1    & 17.14           & \num{4.24e-04}   \\ 
                                        & 14   & 1    & 2.63            & \num{4.24e-04} \\ 
                                        & 28   & 1    & 1.70            & \num{5.00e-04} \\
                                        & 14   & 2    & 1.60            & \num{4.75e-04}  \\ 
                                        & 56   & 1    & 0.98            & \num{4.23e-04} \\
                                        & 14   & 4    & 0.94            & \num{7.61e-04}  \\
                                        & 28   & 2    & 0.81            & \num{5.40e-04} \\ 
                                        & 56   & 2    & 0.67            & \num{4.74e-04}  \\ 
                                        & 56   & 3    & 0.55            & \num{3.16e-04} \\ 
                                        & 56   & 4    & 0.47            & \num{7.60e-04} \\ \bottomrule
    \end{tabular}
    \label{tab:magnon_timings}
\end{table}

As explained in subsection \ref{subsec:sp4_sdc}, although all PFASST runs meet the same residual tolerance, the error generally increases when integrating parallel-in-time. It is also worth noting again here that all field terms except for the exchange field are integrated explicitly when using PFASST. This includes the expensive spin torque evaluations and already roughly halves the runtime compared to BDF on a single core. Lastly, as opposed to the hysteresis simulation in the previous chapter, the error tolerances chosen for this numerical experiment are on the lower end. Higher accuracies are generally desired in this context, where PFASST should perform even better overall.

\section{Conclusions and Discussion}
\label{sec:conclusion}

In this work, we implemented IMEX SDC/PFASST timesteppers for the LLG as extensions to the micromagnetic FEM solver \texttt{magnum.pi}. First numerical experiments lead to the following observations:

IMEX Spectral Deferred Correction method timesteppers are competitive with fully implicit preconditioned BDF(2) methods at moderate accuracies. At higher accuracies, the flexibility of SDC methods enables straightforward construction of high-order IMEX timestepping methods for greatly improved performance.

Parallel-in-time integration of the LLG with decent efficiencies has been achieved across a variety of problem classes (Tab.\ \ref{tab:summary}), where the code generally performs better when high accuracies are desired. Combining space- and time-parallelism leads to better efficiencies than using only space-parallelism --- in particular, parallel-in-time integration is also suited well for parallelism across multiple nodes, where it was successfully used to scale beyond spatial limits.

\begin{table}[]
    \centering\normalsize
    \caption{Maximum speedups $S$ over serial PFASST in the numerical experiments and corresponding parallel efficiencies $\eta=S/n_\text{tot}$, where $n_\text{tot}$ refers to the total number of cores used.}
    \begin{tabular}{lcccccc}
    \toprule
    & \multicolumn{2}{c}{SP4} & \multicolumn{2}{c}{Hysteresis} & \multicolumn{2}{c}{Magnon} \\ \cmidrule(lr){2-3}\cmidrule(lr){4-5}\cmidrule(lr){6-7}
    & $S$ &$\eta$& $S$ &$\eta$& $S$ &$\eta$ \\ \midrule
    Space   & -    &   -    & 5.75 & 0.36 &  17.56 & 0.31        \\ 
    Time    & 2.16 & 0.72 & 3.58 & 0.45& 2.08 & 0.52        \\\midrule
    Total  & 2.16 & 0.72  & 20.56 & 0.16 &  36.51 & 0.16      \\ \bottomrule
    \end{tabular}
    \label{tab:summary}
\end{table}

These preliminary results are best understood as a proof of concept for parallel-in-time integration in micromagnetic simulations, since many optimizations and improvements are yet to be made. For instance, further optimizing the linear and nonlinear solvers leads to faster convergence and enables larger timesteps, which would especially increase performance at low to moderate accuracies. Furthermore, enhancements to SDC such as linearly implicit formulations or variants like inexact SDC \cite{Speck2016} or Krylov-SDC \cite{Huang2006} can yield better convergence. Options for additionally parallelizing SDC ``across the method'' are also available \cite{Speck2018} and the adaptiveness of the algorithm could be improved by increasing/decreasing the number of quadrature nodes automatically as needed. Similarly, adaptive timesteps could be implemented for the PFASST controller. There are also many other parallel-in-time integration algorithms besides PFASST that have yet to be tested in the context of integrating the LLG. Finally, extending GPU simulation codes to multiple GPUs with time parallelism seems promising due to the relatively straightforward implementation. This and the above improvements will be investigated in the future. 

\begin{center}
    \textbf{Acknowledgements}\\
\end{center}

This research was funded in whole, or in part, by the Austrian Science Fund (FWF) P 34671. For the purpose of open access, the author has applied a CC BY public copyright license to any Author Accepted Manuscript version arising from this submission.

\FloatBarrier

\bibliography{main}

\end{document}